\documentclass[12pt]{iopart}
\usepackage{graphicx}
\usepackage{amssymb}
\usepackage{harvard}
\usepackage{soul}

\usepackage{dsfont}
\usepackage[toc,page]{appendix}
\usepackage{braket}
\renewcommand\bra[1]{{\langle{#1}|}}
\renewcommand\ket[1]{{|{#1}\rangle}}
\usepackage{subcaption}
\usepackage{color}
\usepackage{amsthm}
  \expandafter\let\csname equation*\endcsname\relax
\expandafter\let\csname endequation*\endcsname\relax
\usepackage{mathtools}
\usepackage{amsmath}
\usepackage[usenames,dvipsnames]{xcolor}
\usepackage[ocgcolorlinks,colorlinks=true,linkcolor=blue,citecolor=red]{hyperref}
\usepackage{verbatim}
\usepackage{hyperref}
\hypersetup{
	colorlinks=true,
	linkcolor=blue,
	filecolor=magenta,      
	urlcolor=cyan,
}
\usepackage{float}
\usepackage{tcolorbox}
\theoremstyle{definition}

\usepackage{cite}
\DeclareMathOperator{\arcsinh}{arcsinh}

\DeclareMathAlphabet{\mathbbold}{U}{bbold}{m}{n}

\usepackage[nopar]{lipsum}

\makeatletter
\NewEnviron{widerequation}{%
	\begin{equation*}
	\sbox\z@{\let\label\@gobble$\displaystyle\BODY$}
	\makebox[\textwidth]{%
		\begin{minipage}{\dimexpr\wd\z@+3em}
		\vspace{-\baselineskip}
		\begin{equation}
		\BODY
		\end{equation}
		\end{minipage}%
	}
	\end{equation*}
}

\makeatletter
\def\endthebibliography{%
	\def\@noitemerr{\@latex@warning{Empty `thebibliography' environment}}%
	\endlist
}
\makeatother

\begin{document}

\title{Critical heat current for operating an entanglement engine}

\author{Shishir Khandelwal, Nicolas Palazzo, Nicolas Brunner, G\'eraldine Haack}
\address{Department of Applied Physics, University of Geneva, Chemin de Pinchat 22, 1227 Carouge, Gen\`eve, Switzerland}	
\vspace{10pt}

\begin{abstract}
 Autonomous entanglement engines have recently been proposed to generate steady-state bipartite and multipartite entanglement exploiting only incoherent interactions with thermal baths at different temperatures. In this work, we investigate the interplay between heat current and entanglement in a two-qubit entanglement engine, deriving a critical heat current for successful operation of the engine, i.e. a cut-off above which entanglement is present. The heat current can thus be seen as a witness to the presence of entanglement. In the regime of weak-inter-qubit coupling, we also investigate the effect of two experimentally relevant parameters for the qubits, the energy detuning and tunnelling, on the entanglement production. Finally, we show that the regime of strong inter-qubit coupling provides no clear advantage over the weak regime, in the context of out-of-equilibrium entanglement engines.
\end{abstract}

%
%
%
%
%

\section{Introduction}
Autonomous quantum thermal machines - ones that function without external sources of coherence or control, have gathered considerable interest in recent years \cite{Kosloff2014,Mitchison2018a,Mitchison2019}. There are two primary reasons for this interest. Firstly, the elimination of the requirement of classical control may allow one to explore what is truly ``quantum" in quantum thermal machines. Secondly, the thermodynamic cost of classical control can render the thermal machine useless for all practical purposes. Many recent proposals involve autonomous quantum thermal machines performing thermodynamical tasks associated with classical thermal machines such as work extraction and refrigeration \cite{Linden2010,Levy2012,Brunner2012,Roulet2017}, thermometry \cite{Hofer2017a}, clock realization \cite{Erker2017}, and explore how quantum features can be advantageous (see for example \cite{Brunner2014,Maslennikov2019}).  In practice, these ideas can be implemented by exploiting the quantised energy levels in nanostructures \cite{Mari2012,Venturelli2013,BohrBrask2015,Hofer2016,Hofer2016a,Mitchison2016,Josefsson2018}, single-particle quantum coherence \cite{Samuelsson2017,Haack2019}, many-body effects \cite{Chiaracane2020,Doyeux2016,Latune2019} and unconventional materials or phases of matter \cite{Sanchez2015,Sanchez2015a,Roura-Bas2018,Gresta2019}. \par
Remarkably, another type of autonomous thermal machines has recently been put forward, machines that exploit thermal resources to generate entanglement, i.e entanglement engines. They have been explored in the case of bipartite entanglement \cite{BohrBrask2015} and have also recently been proposed to generate multipartite entangled states (for example Bell, GHZ or W-states) \cite{Tavakoli2018, Tavakoli2020} and to generate entanglement using non-thermal baths \cite{Tacchino2018}. Entanglement engines have two key features. Firstly, they provide a simple and exciting platform to study the intersection of quantum information and quantum thermodynamics, which is a growing field of study \cite{thermo1}. Secondly, and importantly, quantum entanglement lacking a classical counterpart, they involve a truly quantum effect in the thermodynamic setting. It is interesting to note that an entanglement engine is not a thermodynamic engine in the usual sense, as there is no work or power involved. However, it is an input-output machine in which heat current (which is a source of free energy) is the input, and the entanglement is the output. In this sense, it is justified to refer to entanglement engines as thermodynamic machines.

   \par

In Ref. \cite{BohrBrask2015}, for the case of weak coupling between qubits, it was shown that a heat current is necessary to generate entanglement in the steady-state. In this work, we establish an exact relation between heat current and entanglement generation, i.e we derive a critical heat current that must be satisfied for the entanglement engine to function successfully. This result can thus be seen as a heat-current-based entanglement witness. In the weak-inter-qubit coupling regime, we also study the effects of detuning (between energy gaps of the qubits) and tunnelling (off-diagonal terms in the Hamiltonians of the qubits); our numerical results show that the qubits can be entangled in the steady-state for a large range of relevant parameters in experiments. The strong-inter-qubit coupling regime, however, requires more careful examination. We show that this case, unlike the weak-coupling case, does not preclude the possibility of having thermal-state entanglement (i.e entanglement at thermal equilibrium). This means that a machine driven by an out-of-equilibrium heat current is no longer necessary to generate entanglement.

\section{Two-qubit entanglement engine in the weak-inter-qubit coupling limit}\label{sec:model}
\begin{figure}[H]
	\centering
	\includegraphics[scale=0.5]{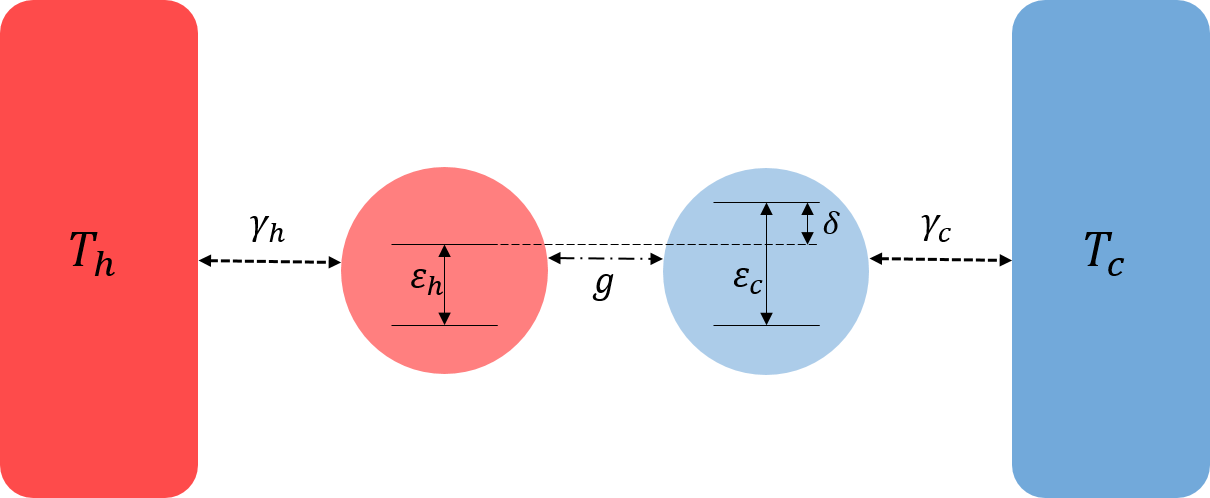}
	\caption{Sketch of an autonomous two-qubit entanglement engine: two qubits are coupled to distinct thermal baths, weakly interacting with each other with strength $g$. Qubit $j$ with energy gap $\varepsilon_j$ is coupled to bath at temperature $T_j$ with strength $\gamma_j$. Energy detuning between the two qubits is labelled $\delta = \varepsilon_c - \varepsilon_h$.}
	\label{fig3}
\end{figure}

We consider two interacting qubits with energy gaps $\varepsilon_h$ and $\varepsilon_c$ (such that $\varepsilon_c-\varepsilon_{h}\coloneqq \delta$) (figure \ref{fig3}) and let them be coupled to distinct thermal baths at temperatures $T_h$ and $T_c$ with $T_h > T_c$, respectively. The qubits are referred to as \textit{hot} and \textit{cold} throughout. Let the Hamiltonian of the system be
\begin{equation}
\begin{aligned}
H &= H_\text{S} + H_{\text{int}} + H_{\text{B}} + H_{\text{SB}}
\end{aligned}
\end{equation}where $H_{\text S}$ is the Hamiltonian of the two qubits, $H_{\text{int}}$ is the interaction between them, $H_{\text B}$ is the Hamiltonian of the two baths and $H_{\text{SB}}$ is the interaction between the qubits and the baths. Explicitly, they take the form:
\begin{equation}\label{eq:ham}
\begin{aligned}
&H_{\text S} = \sum_{j\in\{h,c\}}\varepsilon_j \sigma_+^{(j)}\sigma_-^{(j)}   \quad\quad\quad H_{\text{int}} = g\left(\sigma_+^{(h)}\sigma_-^{(c)}+\sigma_-^{(h)}\sigma_+^{(c)}\right) \\
&H_{\text B} = \sum_{j\in\{h,c\}}\omega_j c_j^{\dagger}c_j \quad \quad \quad\quad\,\, H_{\text{SB}} = \sum_{j\in\{h,c\}}\left(\alpha_j\sigma_-^{(j)}c_j^{\dagger} + \alpha^*_j\sigma_+^{(j)}c_j  \right),
\end{aligned}
\end{equation}with the raising and lowering operators $\sigma_+\coloneqq\ket{1}\!\bra{0}$ and $\sigma_-\coloneqq \ket{0}\!\bra{1}$, with $\ket{1} \coloneqq \left(1,0\right)^T$ and $\ket{0} \coloneqq \left(0,1\right)^T$. The coupling parameters $\alpha_j$s characterise the interaction strength between the qubits with their respective baths. The  operators $c_j$ and $c^{\dagger}_j$ are creation and annihilation operators satisfying the commutation relations that characterise the statistics of the baths. In this work, we restrict to Bosonic baths, for which $[c_i,c^{\dagger}_j] = \mathds{1}\delta_{ij}$.
Born-Markov and  rotating-wave approximations allow the evolution of the two-qubit system to be described by a Lindblad master equation \cite{Breuer2007,Schaller, Potts2019}. To describe the evolution of interacting systems with master equations, one may take the \textit{local} approach in which the environment couples locally with subsystems, or the \textit{global} approach in which the environment couples to global degrees of freedom of the system. The choice depends on the magnitudes of the inter-qubit coupling strength $g$ and the coupling strengths to the baths $\gamma_j$ with respect to the energies of the qubits, and has been discussed at length in \cite{Purkayastha2016,Hofer2017,OnamGonzalez2017,Mitchison2018,Cattaneo2019}. Specifically for the system that we consider, the use of a local master equation is justified if $g\lesssim\gamma_j\ll\varepsilon_j$, and of a global master equation if $g\gg\gamma_{j}$ \cite{Hofer2017}.  Throughout this work, we refer to local (global) and weak-inter-qubit coupling (strong-inter-qubit coupling) interchangeably.\par
 
We first consider the weak-inter-qubit or the local case. In addition to the above mentioned approximations, Markovian dynamics demand that all involved energy scales be much smaller than the energies of the qubits (alternatively, all involved time scales be much larger than the time scales set by the excitation frequencies of the qubits). Since we consider non-degenerate qubits in our analysis, we must impose that the detuning $\delta$ be much smaller than the gaps themselves, $\delta\ll\varepsilon_{j}$.
The evolution of the two qubits is then given by the Lindbladian $\mathcal L$ (note that we set $\hbar=k_B=1$ throughout)
\begin{equation}\label{eq:lind}
\begin{aligned}
\dot\rho(t) &=\mathcal L\rho(t)\\
&= -i[H_{\text{S}}+H_{\text{int}},\rho(t)] + \sum_{j\in\{h,c\}}\gamma_j^+  \mathcal D\left[\sigma_+^{(j)} \right]\rho(t)    +\gamma_j^- \mathcal D\left[\sigma_-^{(j)}  \right]\rho(t)  
\end{aligned}
\end{equation}
with the dissipators $\mathcal{D}\left[A\right]\cdot \coloneqq A\cdot A^{\dagger} - \{A^\dagger A,\cdot\}/2 $, and  the exact coupling rates are determined by the underlying statistics of the baths. Specifically, in the case of coupling to bosonic baths,
\begin{equation}\label{eq:rates}
\gamma_j^+\left(\varepsilon_j\right) = \gamma_j n_B^j\left(\varepsilon_j\right) \quad\text{and}\quad \gamma_j^-\left(\varepsilon_j\right) = \gamma_j \left(1+n^j_B\left(\varepsilon_j\right)\right), \quad j\in\{h,c\}.
\end{equation}The rates $\gamma_{j}$ are defined as $\gamma_{j}\left(E\right) = 2 \pi  \lvert\alpha_j \rvert^2\delta\left(E-\varepsilon_{j}\right)$, where the $\alpha_j$s come from the system-bath Hamiltonian $H_{\text{SB}}$ in ($\ref{eq:ham}$). In this work, we consider energy independent rates of the form $\gamma_j=2\pi\lvert\alpha_j \rvert^2\rho_j$, where $\rho_j$ is a constant density of states in bath $j$ \cite{Schmid1997}. The Bose-Einstein distribution is $n_B^{j} = 1/(e^{\beta_j\varepsilon_j} - 1)$ with the inverse temperature $\beta_j =1/T_j$. As explained above, the dissipative part of the Lindblad equation divides into terms corresponding to hot and cold baths.\par
To solve the Lindblad equation (\ref{eq:lind}) for the steady state, we recast it as a matrix differential equation for the vectorised state $\boldsymbol{p}(t)$ of the density operator $\rho(t)$.

\begin{align}\label{eq:vec}
\rho(t)\longleftrightarrow \boldsymbol{p}(t), \quad \quad \dot{\rho}(t) = \mathcal{L}\rho(t)\longleftrightarrow \dot{\boldsymbol{p}}(t) = M\boldsymbol{p}(t) + \boldsymbol b  
\end{align}
$M$ is a $15\times 15$ matrix and $\boldsymbol b$ is a $15\times 1$ vector; see appendix \ref{app:1} for the exact expressions. 
The steady-state solution $\boldsymbol{p}_{\text{ss}}$ satisfies $\boldsymbol{\dot p_{\text{ss}}}(t)=0$, which upon using (\ref{eq:vec}) gives us $\boldsymbol{p}_{\text{ss}} = -M^{-1}\boldsymbol b$. The form of the interaction Hamiltonian $H_{\text{int}}$ imposes that only two of the off-diagonal elements in the density matrix $\rho(t)$ are non-zero. In the computational basis of the two qubits $\{\ket{11},\ket{10},\ket{01},\ket{00}\}$, the steady-state density matrix, $\rho_{\text{ss}}$ is given by
\begin{align}\label{eq:sstate}
\rho_{\text{ss}} = \begin{pmatrix}
	 r_1 & 0 & 0 & 0 \\
	 0 & r_2 & c & 0 \\
	 0 & c^{*} & r_3 & 0 \\
	 0 & 0 & 0 & r_4 \end{pmatrix}.
\end{align}
The steady-state populations  are given by
\begin{equation}\label{eq:sstate1}
\begin{aligned}
&r_1 =  \frac{4 g^2 (\gamma_{h}^+ +\gamma_{c}^+)^2+\gamma_h^+ \gamma_c^+ \left(\Gamma^2 + 4\delta^2\right)}{\chi } \quad \quad \quad  r_2= \frac{4 g^2 (\gamma_h^- +\gamma_c^-) (\gamma_h^+ +\gamma_c^+)+\gamma_h^+ \gamma_c^- \left(\Gamma^2 + 4\delta^2\right)}{\chi }\\
&r_3 = \frac{4 g^2 (\gamma_h^- +\gamma_c^-) (\gamma_h^+ + \gamma_c^+) + \gamma_h^- \gamma_c^+ \left(\Gamma^2 + 4\delta^2\right)}{\chi } \quad \quad r_4= \frac{4 g^2 (\gamma_{h}^- +\gamma_{c}^-)^2+\gamma_h^- \gamma_c^- \left(\Gamma^2 + 4\delta^2\right)}{\chi }   ,
\end{aligned}
\end{equation}and the coherence (off-diagonal term) $c$ is given by
\begin{align}\label{eq:coh}
c = \frac{2 g   \left(\gamma_h^+ \gamma_c^- - \gamma_h^- \gamma_c^+\right)\left(i\Gamma -2\delta \right)}{\chi }.
\end{align}For convenience, we have introduced the following notation
\begin{equation}
\begin{aligned}
&\chi \coloneqq \left(4 g^2 +\Gamma_h\Gamma_c\right) \Gamma^2 + 4\delta^2\Gamma_h\Gamma_c\\
&\Gamma \coloneqq (\gamma_h^- + \gamma_h^+ + \gamma_c^- + \gamma_c^+)\\
&\Gamma_j \coloneqq \left(\gamma_j^- + \gamma_j^+\right).
\end{aligned}
\end{equation} As expected, in the case where the coupling $g$ between the qubits is set to zero, the coherences vanish, and when there is no detuning between qubits, they are purely imaginary, in agreement with the previous study in\cite{BohrBrask2015}. It is also easy to see that there are no coherences in the reduced steady-states of the two qubits. It is important to note that while the presence of coherence $c$ is essential for the two qubits to be entangled, it is not sufficient to guarantee it. The precise condition on $c$ for the state to be entangled is given in section \ref{sec:crit}.

\section{Steady-state heat current}\label{sec:heat}
\begin{figure}\hspace*{-0.6in}
	\includegraphics[scale=0.7]{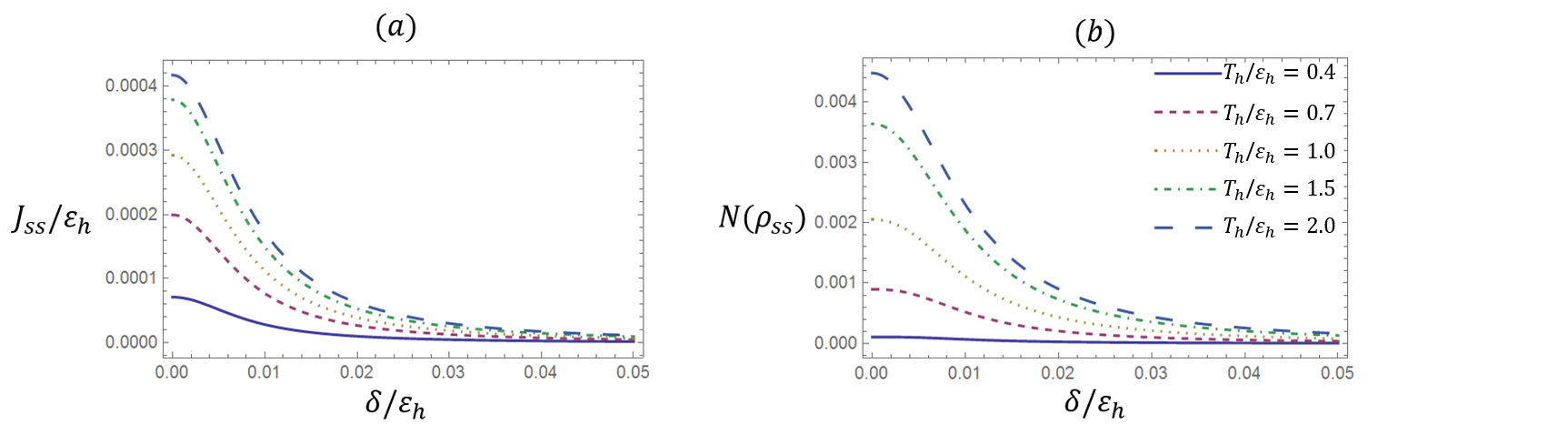}
	\caption{Steady-state (a) heat current $J_{\text{ss}}/\varepsilon_{h}$ and (b) negativity $N(\rho_{\text{ss}})$, as functions of $\delta/\varepsilon_h$, for different values of $T_h/\varepsilon_h$, with $T_c/\varepsilon_h=0.1$, $g/\varepsilon_h=1.6\times10^{-3}$, $\gamma_h/\varepsilon_{h}=10^{-3}$ and $\gamma_c/\varepsilon_{h}=1.1\times10^{-2}$.}\label{fig:heatdet}
\end{figure}

In our master equation approach, the heat flow $Q_j$ from the bath $j$ to qubit $j$ at time $t$ is given by \cite{Alicki1979,Alicki2018}
\begin{align}\label{eq:heatdef}
Q_j(t) = \text{Tr}\left[H_{\text S}\left(\gamma_j^+\mathcal D_j \left[\sigma_+^{\left(j\right)}\right] + \gamma_j^-\mathcal D_j \left[\sigma_-^{\left(j\right)}\right]\right)\rho(t)   \right].
\end{align}Specifically for the steady state, the First Law of thermodynamics \cite{Alicki1979,Alicki2018} imposes that the sum of heat flows from the hot and cold baths must sum to zero, $\sum_{j\in\{c,h\}}Q^{ss}_j = 0$.  Note that we follow the convention in which heat flow from a bath to a qubit is positive, so $Q_c$ takes a negative numerical value, while $Q_h$ takes a positive value. Inserting the steady-state solution (\ref{eq:sstate}), we find the following expression for the steady-state heat current $J_{\text{ss}}$ from the hot bath to the cold bath
\begin{equation} 
\begin{aligned}\label{eq:heatc}
J_{\text{ss}} &\coloneqq Q^{\text{ss}}_h - Q^{\text{ss}}_c\\
&= \frac{8g^2(\gamma_h^+ \gamma_c^- - \gamma_h^- \gamma_c^+)}{\chi}\left(\varepsilon_h\Gamma_c + \varepsilon_c\Gamma_h  \right).
\end{aligned}
\end{equation}The heat current is expected to be of the first-order in the couplings ($\gamma_j$ and $g$). For this to be true, the above mentioned condition $\delta\ll\varepsilon_j$ is essential. Taking the underlying bath (Bose-Einstein) statistics into account in $\gamma_h^- \gamma_c^+ - \gamma_h^+ \gamma_c^-$, we find
\begin{align}
J_{\text{ss}} = \frac{8g^2}{\chi}\gamma_h\gamma_c\left(\varepsilon_h\Gamma_c + \varepsilon_c\Gamma_h \right)\left(n_B^{h}\left(\varepsilon_h\right) - n_B^{c}\left(\varepsilon_{c}\right) \right)\,.
\end{align}The steady-state heat current is proportional to the difference in the distributions of the two baths taken at the energy of each qubit. This expression may remind the reader of the Landauer-B\"uttiker form of the heat current \cite{Moskalets2004,Lesovik2011, Rego1998,Segal2005}, with the energy window for excitations being set by the energy gap of each qubit. Indeed, in the Markovian limit, excitation tunnelling only takes place at resonance (i.e when the excitation frequencies are equal to the energy gaps of the qubits). In figure \ref{fig:heatdet} (a), we show the variation of heat current $J_{\text{ss}}$, with the detuning between energy gaps $\delta$. There is a reduction in heat current for even small values of detuning that we considered. Furthermore, as expected, the heat current increases with increase in $T_h$, with $T_c$ set constant.
\subsection{Heat current and coherence}\label{sec:heatcoh}
Interestingly, there is a similarity in the expressions for the steady-state heat current $J_\text{ss}$ and coherence $c$. Comparing (\ref{eq:coh}) and (\ref{eq:heatc}), we can write
\begin{align}\label{eq:cohj}
c = \frac{J_{\text{ss}}\left(i\Gamma -2\delta\right)}{4g\left(\varepsilon_h\Gamma_c + \varepsilon_c\Gamma_h\right)},
\end{align} which makes it clear that the conditions for heat current and coherence being zero are the same. These conditions are also different from simply having decoupled qubits ($g=0$), in which case the reduced states of the qubits are thermal with the temperatures of their corresponding baths. Therefore the presence of coherence is a signature of the flow of heat and vice-versa.\par

\section{Quantifying steady-state entanglement}\label{sec:entanglement}
 To characterise entanglement, we use the entanglement measure referred to as negativity \cite{Zyczkowski1998,Vidal2002}. For an arbitrary state of two qubits $\rho\in\mathcal S(\mathcal H_h\otimes\mathcal H_c)$, the negativity, $N(\rho)$ is defined as
		\begin{align}
		N(\rho)\coloneqq \sum_{\lambda_i<0}\lvert\lambda_i\rvert,
		\end{align}where $\lambda_i$s are the eigenvalues of the partial transpose of the density matrix  with respect to the one of the qubits. It is simple to see that $N(\rho) = 0$ for a separable state and $N(\rho) = 1/2$ for a Bell state.
	Using the steady-state solution (\ref{eq:sstate}), a straightforward calculation gives
	\begin{align}\label{eq:neg}
	N(\rho_{\text{ss}}) = \text{max}\left\{0,n(\rho_{\text{ss}})\right\},
	\end{align}where
	\begin{align}\label{eq:nrho}
	n(\rho_{\text{ss}}) = \frac{1}{2}\left(\sqrt{4\lvert c\rvert^2 + \left(r_1 - r_4\right)^2 } - \left(r_1 + r_4\right)  \right).
	\end{align}The above equation comes from the fact that only one of the eigenvalues of the partial transpose can be negative in our setup; more details can be found in appendix \ref{app:2}. Upon inserting the steady-state solutions (\ref{eq:sstate1}) and (\ref{eq:coh}) in (\ref{eq:nrho}), we find the following explicit expression for negativity,
	\begin{equation}
	\begin{aligned}
	n(\rho_{\text{ss}}) = \frac{1}{2\chi}\left[-4g^2A -  B\left(4\delta^2+\Gamma^2 \right) + \sqrt{16g^4\left(\gamma_{h}+\gamma_{c}\right)^2\Gamma^2+8g^2C(4\delta^2+\Gamma^2)+D(4\delta^2+\Gamma^2)^2}  \right],
	\end{aligned}
	\end{equation}where
	\begin{equation*}
	A= \left[ \gamma_h^2 + \gamma_c^2 + 2\left(\gamma_h^+ + \gamma_c^- \right) \left(\gamma_h^- + \gamma_c^+ \right)\right],
	\end{equation*}
	\begin{equation*}
	B= \gamma_h^-\gamma_c^- + \gamma_h^+\gamma_c^+,
	\end{equation*}
	\begin{equation*}
	\begin{aligned}
	C =& \gamma_h^+\gamma_c^+\left(\left(\gamma_h^+\right)^2 + \left(\gamma_c^+\right)^2\right) + \gamma_h^-\gamma_c^-\left(\left(\gamma_h^-\right)^2 + \left(\gamma_c^-\right)^2\right) +2\left(\left(\gamma_h^-\right)^2 + \left(\gamma_h^+\right)^2\right)\left(\left(\gamma_c^-\right)^2 + \left(\gamma_c^+\right)^2\right),\\
	& - \left(\gamma_h^-\gamma_h^+ + \gamma_c^-\gamma_c^+\right)\left(\gamma_h^-\gamma_c^+ + \gamma_h^+\gamma_c^-\right) - 8\gamma_h^-\gamma_h^+\gamma_c^-\gamma_c^+,
	\end{aligned}
	\end{equation*}and
	\begin{equation*}
	D=\left(\gamma_h^-\gamma_c^- - \gamma_h^+\gamma_c^+\right)^2.
	\end{equation*}In contrast to the heat current $J_{\text{ss}}$ (\ref{eq:heatc}), the above expressions do not neatly simplify in terms of the difference between the distributions of the two baths. In figure \ref{fig:heatdet} (b), we show the variation of negativity with detuning in the energy gaps. The plots clearly show reduction in the negativity even for a small amount ($5\%$) of non-degeneracy among the qubits. Furthermore, there is an increase in the steady-state negativity with increase in $T_h$. These observations are in direct correspondence with the variation of heat current with detuning that was presented above.
	\subsection{Critical heat current}\label{sec:crit}
     For the presence of non-zero steady-state entanglement between the two qubits, looking at (\ref{eq:nrho}), it is simple to see that the state must satisfy $\lvert c\rvert^2 > r_1r_4$, which is equivalent to the Peres-Horodecki criterion \cite{Peres1996,Horodecki1996}. Interestingly, the populations $r_2$ and $r_3$ do not appear in this bound, despite the presence of coherence in their subspace. One way to see this is that the condition $r_2r_3 > \lvert c\rvert^2$ is satisfied implicitly due to the positivity of the density matrix $\rho_{\text{ss}}$ (which is true because Lindbladian evolution is Completely Positive and Trace Preserving (CPTP)). This condition directly leads us to a lower bound on the heat current required to generate non-zero steady-state entanglement via (\ref{eq:cohj}).
	\begin{align}\label{eq:bound}
	J_{\text{ss}} > 4g\left(\varepsilon_h\Gamma_c + \varepsilon_c\Gamma_h \right)\sqrt{\frac{r_1r_4}{\Gamma^2 + 4\delta^2}  } \coloneqq J_{\text{c}}
	\end{align}The above lower bound can be understood as follows. For given couplings, temperatures and energy gaps in the two-qubit thermal machine, there exists a minimum amount of steady-state heat current required for entanglement to be present. Therefore, a non-zero steady-state heat current is a necessary but not a sufficient condition for entanglement. It leads us to define a \textit{critical} steady-state heat current $J_{\text{c}}$, a cut-off value of heat current needed to form steady-state entanglement between the two qubits. Alternatively, one may regard the quantity $J_{\text{ss}}/J_c$ as a heat-current-based entanglement witness, as the two qubits are entangled in the steady state whenever $J_{\text{ss}}/J_c > 1$. \par
\begin{figure}
 	\centering
 	\includegraphics[scale=0.8]{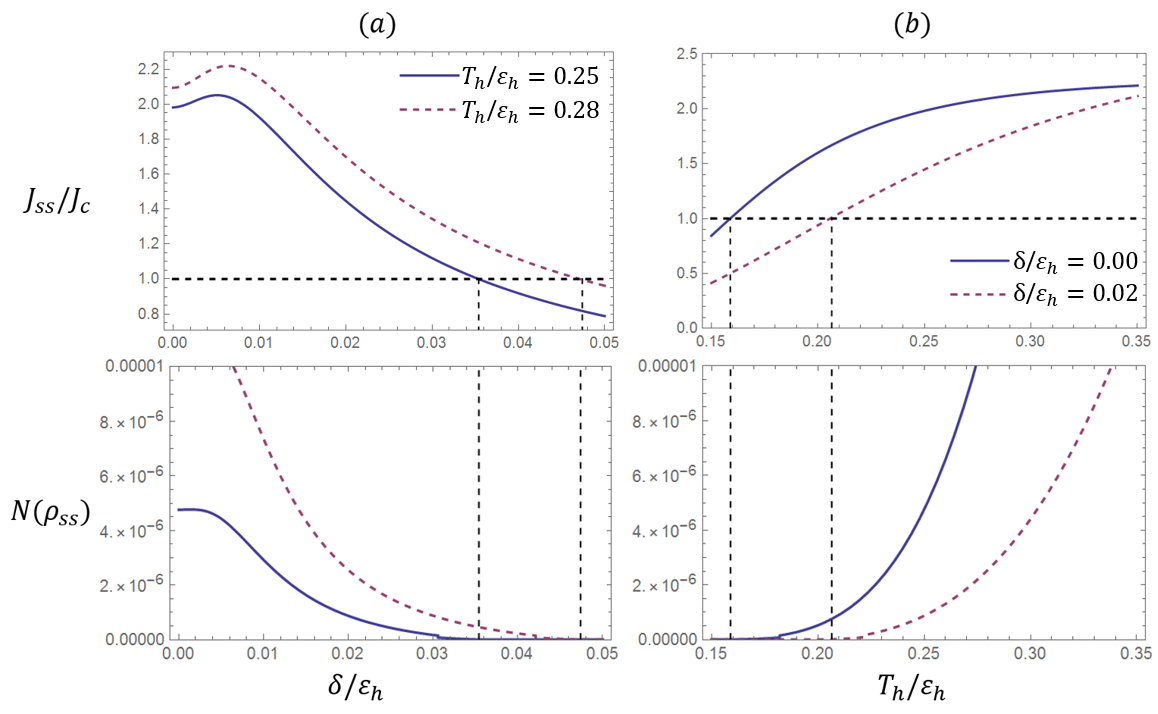}
 	\caption{$J_{\text{ss}}/J_{\text{c}}$ and the negativity, $N(\rho_{\text{ss}})$ as functions of  (a) $\delta/\varepsilon_h$ for constant $T_h/\varepsilon_h$, and (b) $T_h/\varepsilon_h$ for constant $\delta/\varepsilon_h$. $T_c/\varepsilon_h=0.1$, $g/\varepsilon_{h}=1.6\times10^{-3}$, $\gamma_h/\varepsilon_{h}=10^{-3}$ and $\gamma_c/\varepsilon_{h}=1.1\times10^{-2}$.}\label{fig:he}
 \end{figure}

 
 Figure \ref{fig:he} demonstrates the utility of the above analysis. In figure \ref{fig:he} (a), we compare the variation of $J_{\text{ss}}/J_{\text{c}}$ and the steady-state negativity $N(\rho_{\text{ss}})$ with the detuning. We see a small increase and then a monotonic decrease in $J_{\text{ss}}/J_{\text{c}}$ for the considered range of detuning. Observe that the negativity starts from a non-zero value and declines to zero for a particular value of detuning. This is the same value that makes $J_{\text{ss}}/J_{\text{c}} = 1$. Therefore, the lower bound on the heat current allows us to obtain a numerical upper bound on the detuning to have non-zero steady-state entanglement. The analysis is only valid in the regime of small detuning. In figure \ref{fig:he} (b), we perform a similar analysis but with $T_h$ instead of the detuning. In this case, there is a monotonic increase in $J_{\text{ss}}/J_{\text{c}}$. Looking at the variation of the steady-state negativity, we find a numerical lower bound on the hot bath temperature $T_h$ to have non-zero steady-state entanglement.
 
\section{Effect of tunnelling}\label{sec:recovery}
\begin{figure}[H]
		\hspace*{-0.8in}
		\includegraphics[scale=0.7]{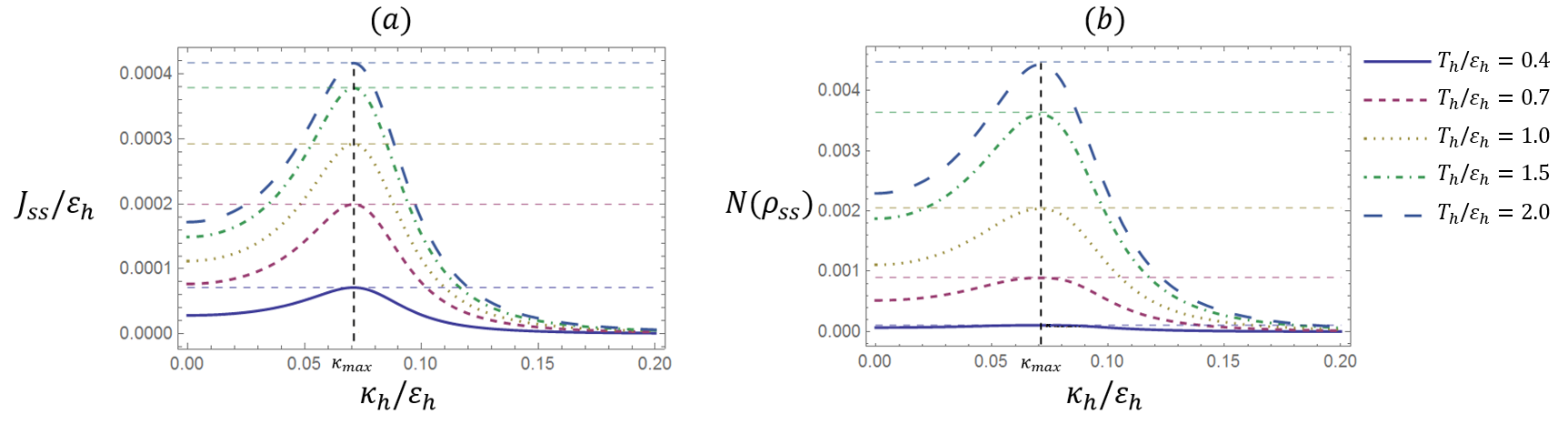}
		\caption{Steady-state heat current $J_{\text{ss}}/\varepsilon_h$ and negativity $N(\rho_{\text{ss}})$ as functions of $\kappa_h/\varepsilon_h$, with $\kappa_c=0$ and $\delta/\varepsilon_h =0.01 $. The horizontal dashed curves mark the heat current and negativity respectively, for $\delta=0$ and $\kappa_h=\kappa_c=0$, for different values of $T_h/\varepsilon_h$. $T_c/\varepsilon_h=0.1$, $g/\varepsilon_{h}=1.6\times10^{-3}$, $\gamma_h/\varepsilon_{h}=10^{-3}$ and $\gamma_c/\varepsilon_{h}=1.1\times10^{-2}$.}
		\label{fig:enttun}
	\end{figure}

	In this section, we numerically analyse the effect of tunnelling terms (off-diagonal elements) in the qubit Hamiltonians on the steady-state heat current and entanglement. We let the Hamiltonian of the system be
	\begin{align}
	H_\text{S} =\sum_{j\in\{h,c\}} \left(\varepsilon_j \sigma_+^{(j)}\sigma_-^{(j)} +\kappa_j\sigma_x^{(j)} \right)  .
	\end{align}In figure \ref{fig:enttun}, we show the variation of the steady-state heat current and negativity, with the tunnelling rate in the hot qubit, $\kappa_h$. Clearly, for smaller magnitudes, the effect of $\kappa_h$ is to counter the effect of detuning; there is an increase in heat current as well as entanglement in the steady-state, with increasing $\kappa_h$. At a certain value of tunnelling rate $\kappa_{\text{max}}$, a maximum value of heat current and entanglement is reached, beyond which the two monotonically decrease. Through our numerical analysis, we find that for $\kappa=\kappa_{\text{max}}$, the heat current and entanglement reach the same values as when $\delta=\kappa_j=0$. It is the tunnelling term that counterbalances the effect of having off-resonant qubits. Interestingly, this maximising value $k_{\max}$ is approximately the same for both heat current and entanglement. Furthermore, there is little variation in $\kappa_{\text{max}}$ with the temperature $T_h$.
\par
\begin{figure}[H]
		\centering
		\includegraphics[scale=0.4]{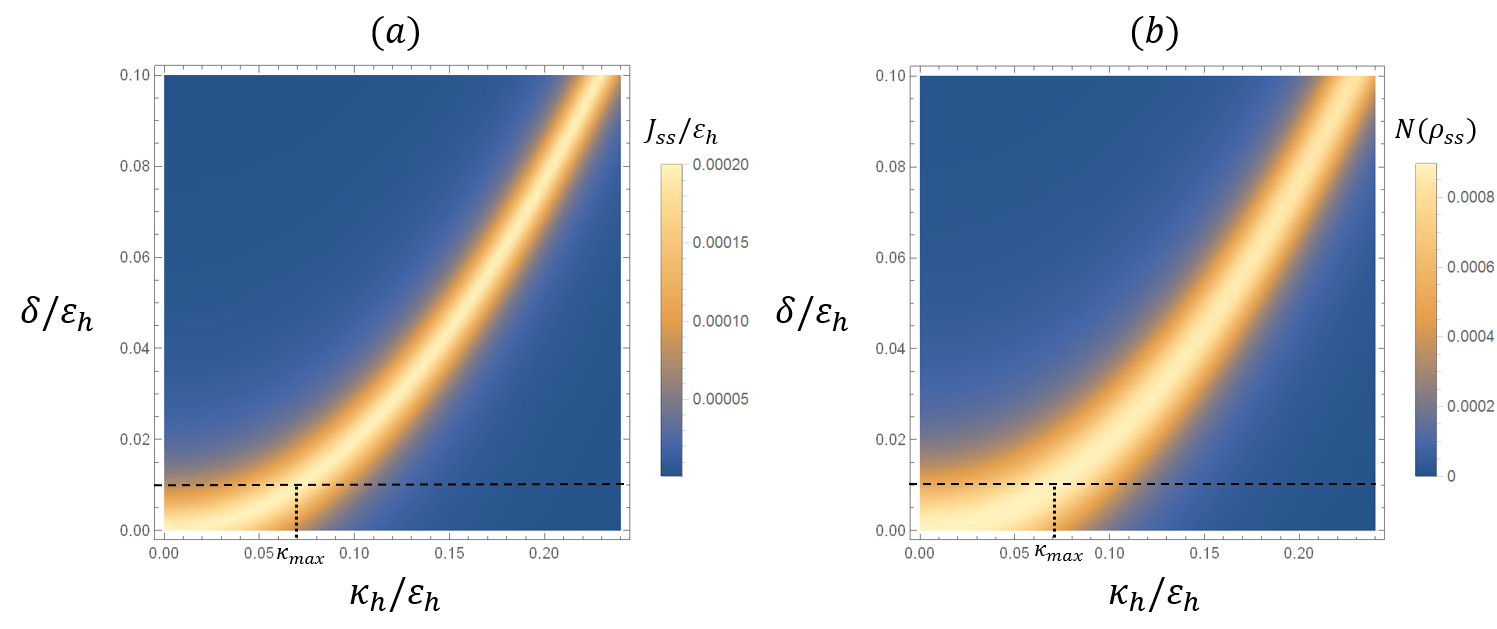}
		\caption{Steady-state (a) heat current $J_{\text{ss}}/\varepsilon_h$ and (b) negativity $N(\rho_{\text{ss}})$ as functions of $\kappa_h/\varepsilon_h$ and $\delta/\varepsilon_h$, with $\kappa_c=0$, $T_h/\varepsilon_h = 0.7$, $T_c/\varepsilon_h = 0.1$, $g/\varepsilon_{h}=1.6\times10^{-3}$, $\gamma_h/\varepsilon_{h}=10^{-3}$ and $\gamma_c/\varepsilon_{h}=1.1\times10^{-2}$. The horizontal dashed lines are representative of the curves corresponding to $T_h/\varepsilon_h=0.7$ in figure \ref{fig:enttun}.}\label{fig:den2}
	\end{figure}

	In figure \ref{fig:den2}, we show the behaviour of steady-state heat current and entanglement respectively with $\kappa_1$ and the detuning $\delta$. The bright region shows the variation of $\kappa_{\text{max}}$ with the detuning. We observe a non-linear but monotonic increase in $\kappa_{\text{max}}$, i.e larger the detuning, greater the magnitude of tunnelling required to recover the heat current and entanglement. The plots show that the two-qubit thermal machine works as an entanglement engine for a large range of parameters, not only restricted to the optimised setting (energy degenerate qubits, no tunnelling) considered in \cite{BohrBrask2015}.
	
	\section{Two-qubit entanglement engine in the strong-inter-qubit coupling limit}

We have studied entanglement production in the two-qubit thermal machine in the limit of weak coupling between the the qubits ($g\lesssim\gamma_{h,c}\ll\varepsilon_{h,c}$). In this section, we consider the case in which the the coupling $g$ is not small compared to the couplings to the baths (but still being less than $\varepsilon_h,\varepsilon_c$); specifically the case in which $g\gg\gamma_{h,c}$. In such a scenario, it is no longer possible to simply decouple the dissipative part in the Lindblad equation into dissipators corresponding to hot and cold baths, and treat the qubit-bath interactions locally. Instead, the baths couple to global degrees of freedom of the two-qubit system, and the Lindblad operators describe the jumps between eigenstates of the total Hamiltonian, $H_{\text{S}}+H_{\text{int}}$ \cite{Correa2013,Hofer2017,Purkayastha2016,Cattaneo2019}. Compared to the local approach, in the dissipative part of the global Lindblad equation, the role of the qubit energies is taken over by the eigenenergies of $H_{\text{S}}+H_{\text{int}}$. \par
Since the main purpose of our analysis is to characterise entanglement creation in the strong-inter-qubit coupling regime, we must mention two caveats concerning this case. The first caveat is that when talking about `entanglement' between systems, a clear bipartition between the systems is assumed. However, in a strong-inter-qubit coupling regime, it is no longer clear whether the two-qubit system should be treated as two interacting subsystems, or as one large system. The second caveat is that the strong-inter-qubit coupling regime does not preclude the possibility of thermal-state entanglement. It can be shown that the thermal state at temperature $T$, $\rho_{\text{th}}=e^{-\left(H_{\text{S}}+H_{\text{int}}\right)/T}/\mathcal Z$ with $\mathcal Z=\text{Tr}\left(e^{-\left(H_{\text{S}}+H_{\text{int}}\right)/T}\right)$, is entangled for $\sinh^2\left(g/T\right)>1$ or $T<g/\arcsinh\left(1\right)$ (see appendix \ref{app:4}). Since we are focused specifically on entanglement creation in the out-of-equilibrium case, it is only appropriate that we try to operate in a regime in which a temperature gradient offers an advantage for entanglement creation. This point will be further discussed below.
\par
For the sake of simplicity and without loss of generality, we restrict to the resonant case in which $\varepsilon_h=\varepsilon_c =\varepsilon$. The eigenenergies of $H_{\text{S}}+H_{\text{int}}$ are $0,\, \varepsilon_-=\varepsilon - g$, $\varepsilon_+=\varepsilon +g$ and $2\varepsilon$, and the corresponding eigenstates are denoted by $\ket{0}$, $\ket{\varepsilon_-}$, $\ket{\varepsilon_+}$ and $\ket{2}$, respectively (see appendix \ref{app:3} for exact expressions). The only non-zero Lindblad operators are the ones corresponding to transitions of energies $\varepsilon_-$ and $\varepsilon_+$. For each of these energies, there are two possible transitions as $\varepsilon_-=\varepsilon_--0=2\varepsilon - \varepsilon_+$ and $\varepsilon_+=\varepsilon_+-0=2\varepsilon-\varepsilon_-$. The Lindblad operators are thus given by \cite{Correa2013,Hofer2017}
	\begin{equation}
	\begin{aligned}
	&\hat L_j(\varepsilon_-) = \ket{0}\!\bra{0}\sigma_-^{\left(j\right)}\ket{\varepsilon_-}\!\bra{\varepsilon_-} + \ket{\varepsilon_+}\!\bra{\varepsilon_+}\sigma_-^{\left(j\right)}\ket{2}\!\bra{2}\\
	&\hat L_j(\varepsilon_+) = \ket{0}\!\bra{0}\sigma_-^{\left(j\right)}\ket{\varepsilon_+}\!\bra{\varepsilon_+} + \ket{\varepsilon_-}\!\bra{\varepsilon_-}\sigma_-^{\left(j\right)}\ket{2}\!\bra{2},
	\end{aligned}
	\end{equation}and the adjoints of the above. The global master equation for the two-qubit entanglement engine then takes the form
	\begin{equation}\label{eq:glind}
	\begin{aligned}
	\dot\rho (t)= -i\left[H_{\text{S}}+H_{\text{int}}, \rho(t)\right]+\sum_{\alpha\in\{-,+\}}\sum_{j\in\{h,c\}}&\gamma_j^{-}\left(\varepsilon_\alpha\right)\mathcal D\left[\hat L_j\left(\varepsilon_\alpha\right)\right]\rho(t)+\gamma_j^{+}\left(\varepsilon_\alpha\right)\mathcal D\left[\hat L^\dagger_j\left(\varepsilon_\alpha\right)\right]\rho(t),
	\end{aligned}
	\end{equation}where the dissipators $\mathcal D\left[A\right]$ were defined in section \ref{sec:model}. In contrast to the local approach, the rates are determined by the eigenenergies $\varepsilon_-$ and $\varepsilon_+$,  $\gamma_j^-\left(\varepsilon_{\pm}\right)=\gamma_j\left(1+n_B^j\left(\varepsilon_{\pm}\right)\right)$ and $\gamma_j^+\left(\varepsilon_{\pm}\right)=\gamma_jn_B^j\left(\varepsilon_{\pm}\right)$ with $j=h,c$. As the form of the inter-qubit interaction remains the same, the form of the steady state for local and global approaches is the same. Using the procedure elaborated in section \ref{sec:model}, we find the following steady-state solution to the Lindblad equation (\ref{eq:glind})
	\begin{align}
	\rho_{\text{ss}}^{\text{gl}} =  \begin{pmatrix}
	s_1 & 0 & 0 & 0 \\
	0 & s_2 & d & 0 \\
	0 & d^{*} & s_3 & 0 \\
	0 & 0 & 0 & s_4 \end{pmatrix}.
	\end{align}The steady-state populations are given by
	\begin{equation}\label{eq:sstate2}
	\begin{aligned}
	&s_1 =  \frac{\Gamma^{+}\left(\varepsilon_-\right) \Gamma^{+}\left(\varepsilon_+\right)}{\chi_{\text{gl}}} \quad \quad \quad \quad\quad\quad s_2= \frac{\Gamma^+\left(\varepsilon_-\right)\Gamma^-\left(\varepsilon_+\right) +\Gamma^-\left(\varepsilon_-\right)\Gamma^+\left(\varepsilon_+\right)}{2\chi_{\text{gl}}}\\
	&s_3= \frac{\Gamma^+\left(\varepsilon_-\right)\Gamma^-\left(\varepsilon_+\right) +\Gamma^-\left(\varepsilon_-\right)\Gamma^+\left(\varepsilon_+\right)}{2\chi_{\text{gl}}}\quad \quad \quad \quad \quad s_4= \frac{\Gamma^{-}\left(\varepsilon_-\right) \Gamma^{-}\left(\varepsilon_+\right)}{\chi_{\text{gl}}},
	\end{aligned}
	\end{equation}and the coherence (off-diagonal term) $d$ is given by
	\begin{align}\label{eq:gcoh}
	d = \frac{\Gamma^{-}\left(\varepsilon_-\right)\Gamma^{+}\left(\varepsilon_+\right) - \Gamma^{+}\left(\varepsilon_-\right)\Gamma^{-}\left(\varepsilon_+\right)}{2\chi_{\text{gl}}},
\end{align}where $\chi_{\text{gl}} = \Gamma\left(\varepsilon_-\right)\Gamma\left(\varepsilon_+\right)$. For convenience, we have defined the sums of rates $\Gamma^{\pm}\left(\varepsilon_{\pm}\right)\coloneqq \gamma^{\pm}_h\left(\varepsilon_{\pm}\right)+\gamma^{\pm}_c\left(\varepsilon_{\pm}\right)$ and $\Gamma\left(\varepsilon_{\pm}\right)=\Gamma^-\left(\varepsilon_{\pm}\right)+\Gamma^+\left(\varepsilon_{\pm}\right)$.

\subsection{Steady-state heat current}
Using the definition of heat current from equation (\ref{eq:heatdef}) and the above steady-state solution (equations (\ref{eq:sstate2}) and (\ref{eq:gcoh})), we find the following expression for steady-state heat current in the global regime
\begin{equation}
\begin{aligned}
J_{\text{ss}}^{\text{gl}} = \frac{\gamma_h\gamma_c}{\chi_{\text{gl}}}\big[&\varepsilon_-\Gamma\left(\varepsilon_+\right)\left(n_B^h\left(\varepsilon_-\right)-n_B^c\left(\varepsilon_-\right)\right)\\
& + \varepsilon_+\Gamma\left(\varepsilon_-\right)\left(n_B^h\left(\varepsilon_+\right) - n_B^c\left(\varepsilon_+\right)\right)\big].
\end{aligned}
\end{equation}The steady-state heat current in the global regime resembles the one obtained for the local regime (equation \ref{eq:heatc}); it is proportional to the the product of qubit-bath couplings $\gamma_h\gamma_c$ and to the difference between distributions of hot and cold baths. A visible distinction lies in the appearance of eigenenergies $\varepsilon_-$ and $\varepsilon_+$. In contrast, in the local approach, in equation (\ref{eq:heatc}), $\varepsilon_h$ and $\varepsilon_c$ were present. When $T_h=T_c$, $n_B^h\left(\varepsilon_{-/+}\right) - n_B^c\left(\varepsilon_{-/+}\right)=0$. Therefore, at thermal equilibrium, we find that $J_{\text{ss}}^{\text{gl}}=0$, as expected.

\subsection{Steady-state entanglement and critical heat current}\label{sec:gcrit} 
\begin{figure}
			\hspace*{-0.7in}
		\centering
		\includegraphics[scale=0.45]{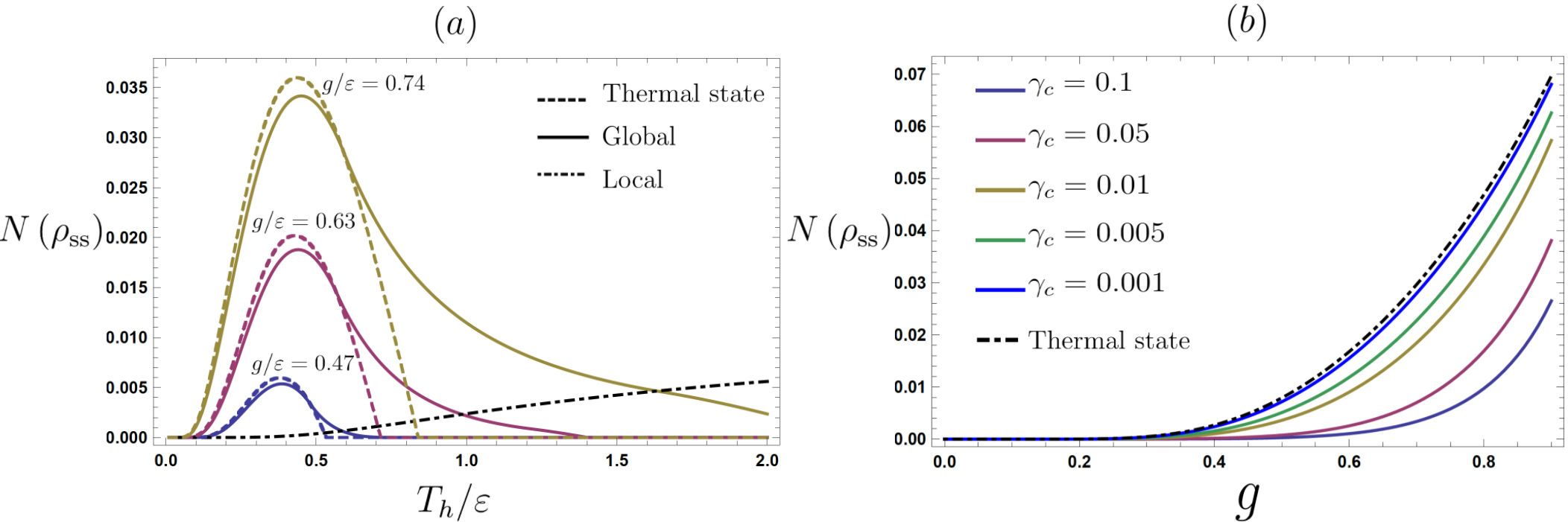}  
	\caption{$(a)$Steady-state  negativity  $N\left(\rho_{\text{ss}}^{\text{gl}}\right)$ as functions of temperature $T_h$, for $T_c/\varepsilon=0.01$, $\gamma_h/\varepsilon=0.01$, $\varepsilon =1$ and different values of $g$. The value of $\gamma_c$ is optimised in the interval $(0.001,0.1)$ to give the maximum negativity. The solid curves correspond to the out-of-equilibrium thermal machine, while the dashed curves correspond to the thermal state with the same value of $g$ and $T=T_h$. Notice that the thermal state negativity falls abruptly to zero when $\sinh\left(g/T\right)\leq1$. The curve for the local case is obtained with parameters $\varepsilon=1$, $\gamma_h/\varepsilon=0.001$, $T_c/\varepsilon=0.01$, and values of $\gamma_c$ and $g$ optimised to maximise negativity. \\$(b)$ $N\left(\rho_{\text{ss}}^{\text{gl}}\right)$ as a function of $g$, with $T_c/\varepsilon=0.01$, $\gamma_h/\varepsilon=0.01$, $\varepsilon =1$ and different values of $\gamma_c$. The value of $T_h$ is optimised to give the maximum negativity. The curve corresponding to the thermal state is obtained by optimising negativity over $T$ with $\varepsilon=1$ and for every value of $g$.}\label{fig:glo}
\end{figure}

As the form of the steady state is the same in both approaches, the form of the steady-state negativity remains the same as in equation (\ref{eq:neg})
\begin{align}
N\left(\rho_{\text{ss}}^{\text{gl}}\right) = \text{max}\left\{0,\frac{1}{2}\left(\sqrt{4\lvert d\rvert^2 + \left(s_1 - s_4\right)^2 } - \left(s_1 + s_4\right)  \right)\right\},
\end{align}where $d$, $s_1$ and $s_4$ are given in equations (\ref{eq:sstate2}) and (\ref{eq:gcoh}). In figure \ref{fig:glo} $(a)$, we plot the negativity as a function of $T_h$, for different values of coupling $g$ and optimised parameter $\gamma_c$. We find that the strong-inter-qubit coupling regime cannot give rise to more entanglement compared to the weak-inter-qubit coupling regime unless the inter-qubit coupling is taken to be comparable to the energy gaps of the qubits, and even then the difference is not significant. It may seem that one can obtain arbitrarily more entanglement by increasing $g$. However, it is important to remember that for our analysis to be meaningful, $g$ must be smaller than $\varepsilon$. A possible way to see this is that if $g=\varepsilon$, two of the eigenvalues of $H_{\text{S}}+H_{\text{int}}$ become zero, and if $g>\varepsilon$, one of the eigenenergies becomes negative and corresponding eigenstate (which is entangled) falls below $\ket{0}$, and becomes the ground state. Therefore, at low temperatures, this entangled eigenstate is largely occupied. Moreover, if $g$ is comparable with  $\varepsilon$, the thermal state with the corresponding temperature can be entangled.  The dashed curves represent thermal state entanglement with the corresponding value of $g$ and $T=T_h$. The plots seem to indicate that the out-of-equilibrium thermal machine is not able to achieve as high entanglement as can be achieved for an equilibrium situation. This is further elaborated in figure \ref{fig:glo} $(b)$, in which we plot the steady-state negativity as a function of $g$ for optimised $T_h$ and different values of $\gamma_c$. We find that as $\gamma_c$ is reduced (i.e the two-qubit system is gradually decoupled from the cold bath), the steady-state entanglement approaches the thermal-state value (the optimal steady-state approaches the thermal state). This shows that the thermal state is in fact optimal for observing entanglement - more entanglement can be observed in a thermal state than in any out-of-equilibrium situation. \par 
The condition for $\rho_{\text{ss}}^{\text{gl}}$ to be entangled is similar to the local case, $\lvert d\rvert^2 >s_1s_4 \implies N\left(\rho_{\text{ss}}^{\text{gl}}\right) >0$. Furthermore, this condition directly leads to a lower bound on heat current, following a similar procedure as in section \ref{sec:crit}. This is given by
\begin{align}
J_{\text{ss}}^{\text{gl}} > \frac{2\gamma_c\gamma_h\left[\varepsilon_-\Gamma\left(\varepsilon_+\right)\left(n_B^h\left(\varepsilon_-\right)-n_B^c\left(\varepsilon_-\right)\right) + \varepsilon_+\Gamma\left(\varepsilon_-\right)\left(n_B^h\left(\varepsilon_+\right) - n_B^c\left(\varepsilon_+\right)\right)\right]}{\Gamma^{-}\left(\varepsilon_-\right)\Gamma^{+}\left(\varepsilon_+\right) - \Gamma^{+}\left(\varepsilon_-\right)\Gamma^{-}\left(\varepsilon_+\right)}\sqrt{s_1s_4}\coloneqq J_{\text{c}}^{\text{gl}}
\end{align}The utility of this lower bound can be understood in the same way as in the local approach (figure \ref{fig:he}); however it does not have the same importance. The steady-state heat current being greater than $J_{\text{c}}^{\text{gl}}$ will indeed guarantee the presence of entanglement. However, as we explained above, in the regime of strong coupling between qubits, the thermal state can be entangled. This means that even in a state of thermal equilibrium (i.e no steady-state heat current), there can be entanglement. In such a case, the critical heat current simply falls to zero. \par
To summarise, the global regime, contrary to na\"ive intuition, does not give rise to considerably more entanglement,, and involves the problem of thermal state entanglement. Moreover, as the plots demonstrated, the entanglement created in the global case is very sensitive to temperature, making it challenging for experiments. Therefore, in the context of exploring entanglement in out-of-equilibrium thermal machines, the local regime offers conceptually and experimentally much more interesting physics.

	\section{Conclusions}
	In the first part of the paper, in the weak-inter-qubit coupling regime, we provided a thorough investigation of the two-qubit thermal machine under a general condition of energy non-degeneracy and the presence of tunnelling. In this case, we operated under a local Lindblad model to obtain a steady-state solution. \par
	Importantly, we established an analytical relation between the steady-state heat current and entanglement generation for our model, through the critical heat current presented in section \ref{sec:crit}. The two-qubit thermal machine only works successfully as an entanglement engine if the temperature gradient is sufficient and the cold bath is at a rather low temperature, as observed in \cite{BohrBrask2015}. This behaviour can now be fully understood through the critical heat current. In addition, we demonstrated that the bound on heat current allows one to determine precise bounds on quantities such as coupling and detuning, in order to see entanglement in the steady-state, i.e to have a working entanglement engine. From a fundamental perspective, the lower bound relates the purely quantum notion of entanglement to the flow of heat.   \par
	Additionally, we presented analytical and numerical results to study the effect of detuning on steady-state heat current and entanglement. Considering the presence of a small detuning, we found a reduction in both the quantities. In the presence of a tunnelling rate within one of the qubits, however, we found that the lost heat current and entanglement can be recovered. Detuning and tunnelling being experimentally relevant parameters, this trade-off holds significance for any future implementations.\par
	Finally, we considered the case of strong coupling between qubits, in which we relied on a global master equation approach. We showed that not only can entanglement exist in the thermal state, more entanglement can be present in this equilibrium state, than for any out-of-equilibrium situation. This means that the regime of strong-inter-qubit coupling is not very interesting from the point of view of out-of-equilibrium thermal machines. \par
	 In this work, we operated exclusively in the steady-state regime. It has been previously shown \cite{Brask2015a,Mitchison2015a} that the transient regime can be advantageous for reaching a larger amount of entanglement. Hence, an open question concerns the behaviour of the critical current in the transient case. Furthermore, it would also be of interest to go beyond the two-qubit model to multipartite engines \cite{Tavakoli2020}, to see the interplay between heat flow and entanglement generation.

\section*{Acknowledgements} 
We thank Jonatan Bohr Brask for providing feedback on the manuscript. We acknowledge support from the Swiss National Science Foundation through the starting grant PRIMA  PR00P2$\_$179748, the grant 200021$\_$169002, and through the NCCR QSIT - Quantum Science and Technology. 

\section*{References}

\bibliography{library}

\providecommand{\newblock}{}
\begin{thebibliography}{10}
\expandafter\ifx\csname url\endcsname\relax
  \def\url#1{{\tt #1}}\fi
\expandafter\ifx\csname urlprefix\endcsname\relax\def\urlprefix{URL }\fi
\providecommand{\eprint}[2][]{\url{#2}}

\bibitem{Kosloff2014}
Kosloff R and Levy A 2014 {\em Ann. Rev. Phys. Chem.\/} {\bf 65} 365--393
  \urlprefix\url{https://www.annualreviews.org/doi/abs/10.1146/annurev-physchem-040513-103724}

\bibitem{Mitchison2018a}
Mitchison M~T and Potts P~P 2018 {Physical Implementations of Quantum
  Absorption Refrigerators} {\em Thermodynamics in the Quantum Regime\/}
  (Springer, Cham) chap~6, pp 149--174
  \urlprefix\url{https://link.springer.com/chapter/10.1007/978-3-319-99046-0{\_}6}

\bibitem{Mitchison2019}
Mitchison M~T 2019 {\em Contemp. Phys.\/} {\bf 60} 164--187
  \urlprefix\url{https://www.tandfonline.com/doi/full/10.1080/00107514.2019.1631555}

\bibitem{Linden2010}
Linden N, Popescu S and Skrzypczyk P 2010 {\em Phys. Rev. Lett.\/} {\bf 105}
  130401
  \urlprefix\url{https://journals.aps.org/prl/abstract/10.1103/PhysRevLett.105.130401}

\bibitem{Levy2012}
Levy A and Kosloff R 2012 {\em Phys. Rev. Lett.\/} {\bf 108} 070604
  \urlprefix\url{https://journals.aps.org/prl/abstract/10.1103/PhysRevLett.108.070604}

\bibitem{Brunner2012}
Brunner N, Linden N, Popescu S and Skrzypczyk P 2012 {\em Phys. Rev. E\/} {\bf
  85} 051117
  \urlprefix\url{https://journals.aps.org/pre/abstract/10.1103/PhysRevE.85.051117}

\bibitem{Roulet2017}
Roulet A, Nimmrichter S, Arrazola J~M, Seah S and Scarani V 2017 {\em Phys.
  Rev. E\/} {\bf 95} 062131
  \urlprefix\url{https://journals.aps.org/pre/abstract/10.1103/PhysRevE.95.062131}

\bibitem{Hofer2017a}
Hofer P~P, Brask J~B, Perarnau-Llobet M and Brunner N 2017 {\em Phys. Rev.
  Lett.\/} {\bf 119} 090603
  \urlprefix\url{https://journals.aps.org/prl/abstract/10.1103/PhysRevLett.119.090603}

\bibitem{Erker2017}
Erker P, Mitchison M~T, Silva R, Woods M~P, Brunner N and Huber M 2017 {\em
  Phys. Rev. X\/} {\bf 7} 031022
  \urlprefix\url{http://link.aps.org/doi/10.1103/PhysRevX.7.031022}

\bibitem{Brunner2014}
Brunner N, Huber M, Linden N, Popescu S, Silva R and Skrzypczyk P 2014 {\em
  Phys. Rev. E\/} {\bf 89}
  \urlprefix\url{https://journals.aps.org/pre/abstract/10.1103/PhysRevE.89.032115}

\bibitem{Maslennikov2019}
Maslennikov G, Ding S, Habl{\"{u}}tzel R, Gan J, Roulet A, Nimmrichter S, Dai
  J, Scarani V and Matsukevich D 2019 {\em Nat. Commun.\/} {\bf 10} 202
  \urlprefix\url{https://www.nature.com/articles/s41467-018-08090-0{\#}citeas}

\bibitem{Mari2012}
Mari A and Eisert J 2012 {\em Phys. Rev. Lett.\/} {\bf 108} 120602
  \urlprefix\url{https://journals.aps.org/prl/abstract/10.1103/PhysRevLett.108.120602}

\bibitem{Venturelli2013}
Venturelli D, Fazio R and Giovannetti V 2013 {\em Phys. Rev. Lett.\/} {\bf 110}
  256801
  \urlprefix\url{https://journals.aps.org/prl/abstract/10.1103/PhysRevLett.110.256801}

\bibitem{BohrBrask2015}
{Bohr Brask} J, Haack G, Brunner N and Huber M 2015 {\em New J. Phys.\/} {\bf
  17} 113029
  \urlprefix\url{https://iopscience.iop.org/article/10.1088/1367-2630/17/11/113029}

\bibitem{Hofer2016}
Hofer P~P, Perarnau-Llobet M, Brask J~B, Silva R, Huber M and Brunner N 2016
  {\em Phys. Rev. B\/} {\bf 94} 235420
  \urlprefix\url{https://journals.aps.org/prb/abstract/10.1103/PhysRevB.94.235420}

\bibitem{Hofer2016a}
Hofer P~P, Souquet J~R and Clerk A~A 2016 {\em Phys. Rev. B\/} {\bf 93} 041418
  \urlprefix\url{https://journals.aps.org/prb/abstract/10.1103/PhysRevB.93.041418}

\bibitem{Mitchison2016}
Mitchison M~T, Huber M, Prior J, Woods M~P and Plenio M~B 2016 {\em Quant. Sci.
  Tech.\/} {\bf 1} 15001
  \urlprefix\url{https://iopscience.iop.org/article/10.1088/2058-9565/1/1/015001/meta}

\bibitem{Josefsson2018}
Josefsson M, Svilans A, Burke A~M, Hoffmann E~A, Fahlvik S, Thelander C,
  Leijnse M and Linke H 2018 {\em Nat. Nanotech.\/} {\bf 13} 920--924
  \urlprefix\url{https://www.nature.com/articles/s41565-018-0200-5}

\bibitem{Samuelsson2017}
Samuelsson P, Kheradsoud S and Sothmann B 2017 {\em Phys. Rev. Lett.\/} {\bf
  118} 256801
  \urlprefix\url{https://journals.aps.org/prl/abstract/10.1103/PhysRevLett.118.256801}

\bibitem{Haack2019}
Haack G and Giazotto F 2019 {\em Phys. Rev. B\/} {\bf 100} 235442
  \urlprefix\url{https://journals.aps.org/prb/abstract/10.1103/PhysRevB.100.235442}

\bibitem{Chiaracane2020}
Chiaracane C, Mitchison M~T, Purkayastha A, Haack G and Goold J 2020 {\em Phys.
  Rev. Res.\/} {\bf 2} 013093
  \urlprefix\url{https://journals.aps.org/prresearch/abstract/10.1103/PhysRevResearch.2.013093}

\bibitem{Doyeux2016}
Doyeux P, Leggio B, Messina R and Antezza M 2016 {\em Phys. Rev. E\/} {\bf 93}
  022134
  \urlprefix\url{https://journals.aps.org/pre/abstract/10.1103/PhysRevE.93.022134}

\bibitem{Latune2019}
Latune C~L, Sinayskiy I and Petruccione F 2019 {\em Sci. Rep.\/} {\bf 9} 3191
  \urlprefix\url{https://www.nature.com/articles/s41598-019-39300-4}

\bibitem{Sanchez2015}
S{\'{a}}nchez R, Sothmann B and Jordan A~N 2015 {\em Phys. Rev. Lett.\/} {\bf
  114} 146801
  \urlprefix\url{https://journals.aps.org/prl/abstract/10.1103/PhysRevLett.114.146801}

\bibitem{Sanchez2015a}
S{\'{a}}nchez R, Sothmann B and Jordan A~N 2015 {\em New J. Phys.\/} {\bf 17}
  075006
  \urlprefix\url{https://iopscience.iop.org/article/10.1088/1367-2630/17/7/075006/meta}

\bibitem{Roura-Bas2018}
Roura-Bas P, Arrachea L and Fradkin E 2018 {\em Phys. Rev. B\/} {\bf 98} 195429
  \urlprefix\url{https://journals.aps.org/prb/abstract/10.1103/PhysRevB.98.195429}

\bibitem{Gresta2019}
Gresta D, Real M and Arrachea L 2019 {\em Phys. Rev. Lett.\/} {\bf 123} 186801
  \urlprefix\url{https://journals.aps.org/prl/abstract/10.1103/PhysRevLett.123.186801}

\bibitem{Tavakoli2018}
Tavakoli A, Haack G, Huber M, Brunner N and Brask J~B 2018 {\em Quantum\/} {\bf
  2} 73 \urlprefix\url{https://quantum-journal.org/papers/q-2018-06-13-73/}

\bibitem{Tavakoli2020}
Tavakoli A, Haack G, Brunner N and Brask J~B 2020 {\em Phys. Rev. A\/} {\bf
  101} 012315
  \urlprefix\url{https://link.aps.org/doi/10.1103/PhysRevA.101.012315}

\bibitem{Tacchino2018}
Tacchino F, Auff\`eves A, Santos M~F and Gerace D 2018 {\em Phys. Rev. Lett.\/}
  {\bf 120} 063604
  \urlprefix\url{https://link.aps.org/doi/10.1103/PhysRevLett.120.063604}

\bibitem{thermo1}
Goold J, Huber M, Riera A, {Del Rio} L and Skrzypczyk P 2016 {\em J. Phys. A\/}
  {\bf 49} 14
  \urlprefix\url{https://iopscience.iop.org/article/10.1088/1751-8113/49/14/143001/meta}

\bibitem{Breuer2007}
Breuer H~P and Petruccione F 2007 {\em {The Theory of Open Quantum Systems}\/}
  vol~1 (Oxford University Press)

\bibitem{Schaller}
Schaller G 2014 {\em {Open Quantum Systems Far from Equilibrium}\/} (Springer,
  Cham)

\bibitem{Potts2019}
Potts P~P 2019  \urlprefix\url{http://arxiv.org/abs/1906.07439}

\bibitem{Purkayastha2016}
Purkayastha A, Dhar A and Kulkarni M 2016 {\em Phys. Rev. A\/} {\bf 93} 062114
  \urlprefix\url{https://link.aps.org/doi/10.1103/PhysRevA.93.062114}

\bibitem{Hofer2017}
Hofer P~P, Perarnau-Llobet M, Miranda L~D~M, Haack G, Silva R, Brask J~B and
  Brunner N 2017 {\em New J. Phys.\/} {\bf 19} 123037
  \urlprefix\url{https://iopscience.iop.org/article/10.1088/1367-2630/aa964f}

\bibitem{OnamGonzalez2017}
{Onam Gonz{\'{a}}lez} J, Correa L~A, Nocerino G, Palao J~P, Alonso D and Adesso
  G 2017 {\em Open Sys. Inf. Dyn.\/} {\bf 24} 1740010
  \urlprefix\url{https://www.worldscientific.com/doi/abs/10.1142/S1230161217400108}

\bibitem{Mitchison2018}
Mitchison M~T and Plenio M~B 2018 {\em New J. Phys.\/} {\bf 20} 033005
  \urlprefix\url{https://iopscience.iop.org/article/10.1088/1367-2630/aa9f70}

\bibitem{Cattaneo2019}
Cattaneo M, Giorgi G~L, Maniscalco S and Zambrini R 2019 {\em New J. Phys.\/}
  {\bf 21} 113045 ISSN 1367-2630
  \urlprefix\url{https://iopscience.iop.org/article/10.1088/1367-2630/ab54ac}

\bibitem{Schmid1997}
Schmid J, K{\"{o}}nig J, Schoeller H and Sch{\"{o}}n G 1997 {\em Physica E\/}
  {\bf 1} 241--244
  \urlprefix\url{https://www.sciencedirect.com/science/article/pii/S1386947797000519}

\bibitem{Alicki1979}
Alicki R 1979 {\em J. Phys. A\/} {\bf 12} 5
  \urlprefix\url{https://iopscience.iop.org/article/10.1088/0305-4470/12/5/007/meta}

\bibitem{Alicki2018}
Alicki R and Kosloff R 2018 {Introduction to Quantum Thermodynamics: History
  and Prospects} {\em Thermodynamics in the Quantum Regime\/} ed Binder F,
  Correa L~A, Gogolin C, Anders J and Adesso G (Springer, Cham) pp 1--33
  \urlprefix\url{https://link.springer.com/chapter/10.1007/978-3-319-99046-0{\_}1}

\bibitem{Moskalets2004}
Moskalets M and B{\"{u}}ttiker M 2004 {\em Phys. Rev. B\/} {\bf 70} 245305
  \urlprefix\url{https://journals.aps.org/prb/abstract/10.1103/PhysRevB.70.245305}

\bibitem{Lesovik2011}
Lesovik G~B and Sadovskyy I~A 2011 {\em Phys.-Uspekhi\/} {\bf 54} 1007
  \urlprefix\url{https://iopscience.iop.org/article/10.3367/UFNe.0181.201110b.1041}

\bibitem{Rego1998}
Rego L~G and Kirczenow G 1998 {\em Phys. Rev. Lett.\/} {\bf 81} 232
  \urlprefix\url{https://journals.aps.org/prl/abstract/10.1103/PhysRevLett.81.232}

\bibitem{Segal2005}
Segal D and Nitzan A 2005 {\em Phys. Rev. Lett.\/} {\bf 94} 034301
  \urlprefix\url{https://link.aps.org/doi/10.1103/PhysRevLett.94.034301}

\bibitem{Zyczkowski1998}
{\.{Z}}yczkowski K, Horodecki P, Sanpera A and Lewenstein M 1998 {\em Phys.
  Rev. A\/} {\bf 58} 883--892
  \urlprefix\url{https://link.aps.org/doi/10.1103/PhysRevA.58.883}

\bibitem{Vidal2002}
Vidal G and Werner R~F 2002 {\em Phys. Rev. A\/} {\bf 65} 032314
  \urlprefix\url{https://journals.aps.org/pra/abstract/10.1103/PhysRevA.65.032314}

\bibitem{Peres1996}
Peres A 1996 {\em Phys. Rev. Lett.\/} {\bf 77} 1413--1415
  \urlprefix\url{https://journals.aps.org/prl/abstract/10.1103/PhysRevLett.77.1413}

\bibitem{Horodecki1996}
Horodecki M, Horodecki P and Horodecki R 1996 {\em Phys. Lett. A\/} {\bf
  223:1-2} 1--8
  \urlprefix\url{https://www.sciencedirect.com/science/article/pii/S0375960196007062}

\bibitem{Correa2013}
Correa L~A, Palao J~P, Adesso G and Alonso D 2013 {\em Phys. Rev. E\/} {\bf 87}
  042131
  \urlprefix\url{https://journals.aps.org/pre/abstract/10.1103/PhysRevE.87.042131}

\bibitem{Brask2015a}
Brask J~B and Brunner N 2015 {\em Phys. Rev. E\/} {\bf 92} 062101
  \urlprefix\url{https://journals.aps.org/pre/abstract/10.1103/PhysRevE.92.062101}

\bibitem{Mitchison2015a}
Mitchison M~T, Woods M~P, Prior J and Huber M 2015 {\em New J. Phys.\/} {\bf
  17} 115013
  \urlprefix\url{https://iopscience.iop.org/article/10.1088/1367-2630/17/11/115013}

\end{thebibliography}

\begin{appendices}
\section{Weak-inter-qubit coupling regime}
	\subsection{Steady-state solution}\label{app:1}
	To obtain the steady-state solution in section \ref{sec:model}, we recast (\ref{eq:lind}) as a matrix differential equation for the vectorised state $\boldsymbol{p}(t)$ of the density operator $\rho(t)$.
	\begin{align}\label{eq:vec}
	\rho(t)\longleftrightarrow \boldsymbol{p}(t), \quad \quad \dot{\rho}(t) = \mathcal{L}\rho(t)\longleftrightarrow \dot{\boldsymbol{p}}(t) = M\boldsymbol{p}(t) + \boldsymbol b  
	\end{align}$M$ is a $15\times 15$ matrix, and $\boldsymbol p$ and $\boldsymbol b$ are $15\times 1$ vectors (one element of $\boldsymbol p$ can be eliminated using $\text{Tr}\left(\rho(t)\right)=1$), which are explicitly given by
	\begin{align*}
	M=
	\end{align*}
	\begin{widerequation}
	\resizebox{1.1\hsize}{!}{$\left(
	\begin{array}{ccccccccccccccc}
	-\gamma_h^- - \gamma_c^- & 0 & 0 & 0 & 0 & \gamma_c^+ & 0 & 0 & 0 & 0 & \gamma_h^+ & 0 & 0 & 0 & 0 \\
	0 & \frac{1}{2} (-2 \gamma_h^- -\Gamma_c)-i \varepsilon _c & i g & 0 & 0 & 0 & 0 & 0 & 0 & 0 & 0 & \gamma_h^+ & 0 & 0 & 0 \\
	0 & i g & \frac{1}{2} (-\Gamma_h - 2\gamma_c^-) - i \varepsilon _h & 0 & 0 & 0 & 0 & \gamma_c^+ & 0 & 0 & 0 & 0 & 0 & 0 & 0 \\
	0 & 0 & 0 & -\frac{1}{2}\Gamma-i \left(\varepsilon _h+\varepsilon _c\right) & 0 & 0 & 0 & 0 & 0 & 0 & 0 & 0 & 0 & 0 & 0 \\
	0 & 0 & 0 & 0 & \frac{1}{2} (-2 \gamma_h^- -\Gamma_c)+i \varepsilon _c & 0 & 0 & 0 & -i g & 0 & 0 & 0 & 0 & 0 & \gamma_h^+ \\
	\gamma_c^- -\gamma_h^+ & 0 & 0 & 0 & 0 & -\Gamma_h-\gamma_c^+ & i g & 0 & 0 & -i g & -\gamma_h^+ & 0 & 0 & 0 & 0 \\
	0 & 0 & 0 & 0 & 0 & i g & -\frac{1}{2}\Gamma + i \delta & 0 & 0 & 0 & -i g & 0 & 0 & 0 & 0 \\
	0 & 0 & \gamma_c^- & 0 & 0 & 0 & 0 & \frac{1}{2} (-\Gamma_h - 2 \gamma_c^+)-i \varepsilon_h & 0 & 0 & 0 & -i g & 0 & 0 & 0 \\
	0 & 0 & 0 & 0 & -i g & 0 & 0 & 0 & \frac{1}{2} (-\Gamma_h - 2 \gamma_c^-) + i \varepsilon _h & 0 & 0 & 0 & 0 & \gamma_c^+ & 0 \\
	0 & 0 & 0 & 0 & 0 & -i g & 0 & 0 & 0 & -\frac{1}{2} \Gamma - i \delta & i g & 0 & 0 & 0 & 0 \\
	\gamma_h^- - \gamma_c^+ & 0 & 0 & 0 & 0 & -\gamma_c^+ & -i g & 0 & 0 & i g & -\gamma_h^+ - \Gamma_c & 0 & 0 & 0 & 0 \\
	0 & \gamma_h^- & 0 & 0 & 0 & 0 & 0 & -i g & 0 & 0 & 0 & \frac{1}{2} (-2 \gamma_h^+ -\Gamma_c)-i \varepsilon _c & 0 & 0 & 0 \\
	0 & 0 & 0 & 0 & 0 & 0 & 0 & 0 & 0 & 0 & 0 & 0 & -\frac{1}{2}\Gamma + i \left(\varepsilon _h+\varepsilon_c\right) & 0 & 0 \\
	0 & 0 & 0 & 0 & 0 & 0 & 0 & 0 & \gamma_c^- & 0 & 0 & 0 & 0 & \frac{1}{2} (-\Gamma_h - 2 \gamma_c^+)+i \varepsilon_h & i g \\
	0 & 0 & 0 & 0 & \gamma_h^- & 0 & 0 & 0 & 0 & 0 & 0 & 0 & 0 & i g & \frac{1}{2} (-2 \gamma_h^+ - \Gamma_c)+i \varepsilon_c \\
	\end{array}
	\right)$}
	\end{widerequation}and
	\begin{equation}
	\boldsymbol{b}=\left(0,0,0,0,0,\gamma_h^+,0,0,0,0,\gamma_c^+,0,0,0,0 \right)^T.
	\end{equation}The steady-state solution is simply given by $\boldsymbol{p_{\text{ss}}} = -M^{-1}\boldsymbol b$.
	
	\subsection{Steady-state negativity}\label{app:2}
	We know that the form of the interaction Hamiltonian $H_{\text{int}}$ imposes that the steady-state of the two qubits takes the form given by (\ref{eq:sstate}). The eigenvalues $\lambda_j$ of its partial transpose are listed below.
	\begin{equation}
	\begin{aligned}
	&\lambda_1 = r_2 \quad\quad \lambda_3 =\frac{1}{2} \left(-\sqrt{4\lvert c\rvert^2+(r_1-r_4)^2} + r_1 + r_4\right)\\
	&\lambda_2 = r_3 \quad\quad \lambda_4 =\frac{1}{2} \left(\sqrt{4\lvert c\rvert^2+(r_1-r_4)^2} + r_1 + r_4\right)
	\end{aligned}
	\end{equation}Since only $\lambda_3$ can take a negative value, the steady-state negativity is simply given by
	\begin{align}
	N(\rho_{\text{ss}}) = \text{max}\left\{0,-\lambda_3\right\}.
	\end{align}
	
	\section{Strong-inter-qubit coupling regime}
	\subsection{Eigenstates of the Hamiltonian}\label{app:3}
	The eigenstates of $H_{\text{S}}+H_{\text{int}}$ are given by
	\begin{equation}
	\begin{aligned}
&\ket{0} =\ket{00} \quad\quad\quad\quad\quad\quad\quad\quad \ket{\varepsilon_-}=\frac{1}{\sqrt{2}}\left(\ket{01}-\ket{10}\right)\\
&\ket{\varepsilon_+}=\frac{1}{\sqrt{2}}\left(\ket{01}+\ket{10}\right)\quad\quad \ket{2}=\ket{11}.
	\end{aligned}
	\end{equation}Note that the eigenstates $\ket{\varepsilon_{\pm}}$ are entangled.

	\subsection{Thermal state of two qubits}\label{app:4}
	The thermal state of two qubits with total Hamiltonian $H_{\text{S}}+H_{\text{int}}$ at temperature $T$ is given by
	\begin{equation}
	\begin{aligned}
	\rho_{\text{th}}&=\frac{e^{-\left(H_{\text{S}}+H_{\text{int}}\right)/T}}{\text{Tr}\left(e^{-\left(H_{\text{S}}+H_{\text{int}}\right)/T}\right)}\\
	&=\frac{1}{2 \left(\cosh \left(\frac{\varepsilon}{T}\right)+\cosh \left(\frac{g}{T}\right)\right)}\left(
	\begin{array}{cccc}
	e^{-\varepsilon/T} & 0 & 0 & 0 \\
	0 & \cosh\left(\frac{g}{T}\right) & -\sinh \left(\frac{g}{T}\right) & 0 \\
	0 & -\sinh \left(\frac{g}{T}\right) & \cosh\left(\frac{g}{T}\right) & 0 \\
	0 & 0 & 0 & e^{\varepsilon/T} \\
	\end{array}
	\right),
	\end{aligned}
\end{equation}and the thermal-state negativity is given by
	\begin{align}
	N\left(\rho_{\text{th}}\right)= \text{max}\left\{0,\frac{\sqrt{\cosh^2\left(\frac{\varepsilon}{T}\right)+\sinh^2\left(\frac{g}{T}\right)-1}-\cosh\left(\frac{\varepsilon}{T}\right)}{2 \left(\cosh \left(\frac{\varepsilon}{T}\right)+\cosh \left(\frac{g}{T}\right)\right)}\right\}.
	\end{align}Clearly, $N\left(\rho_{\text{th}}\right)>0\iff \sinh^2\left(g/T\right)>1$.
	\end{appendices}

\end{document}